\documentclass{article}
\pdfoutput=1

\usepackage{PRIMEarxiv}

\usepackage[utf8]{inputenc} 
\usepackage[T1]{fontenc}    
\usepackage{hyperref}       
\usepackage{url}            
\usepackage{booktabs}       
\usepackage{amsfonts}       
\usepackage{nicefrac}       
\usepackage{microtype}      
\usepackage{lipsum}
\usepackage{fancyhdr}       
\usepackage{graphicx}       
\usepackage{amsmath}
\usepackage{subfig}

\usepackage{amssymb,amsbsy,amsthm,booktabs,graphicx,rotating}

\title{A Statistical Model of Serve Return Impact Patterns in Professional Tennis}

\author{Stephanie A. Kovalchik \\
	Zelus Analytics \\
	Austin, Texas \\
	\texttt{skovalchik@zelusanalytics.com} \\
	\And
	Jim Albert \\
	Bowling Green State University\\ 
	Bowling Green, Ohio \\
	\texttt{albert@bgsu.edu} \\
}

\renewcommand\toprule{\\[-5.5pt]\hline\\[-5pt]}
\renewcommand\midrule{\\[-7.5pt]\hline\\[-5pt]}
\renewcommand\bottomrule{\\[-7.5pt]\hline}

\begin{document}

\maketitle

\begin{abstract}
The spread in the use of tracking systems in sport has made fine-grained spatiotemporal analysis a primary focus of an emerging sports analytics industry. Recently publicized tracking data for men's professional tennis allows for the first detailed spatial analysis of return impact. Mixture models are an appealing model-based framework for spatial analysis in sport, where latent variable discovery is often of primary interest. Although finite mixture models have the advantages of interpretability and scalability, most implementations assume standard parametric distributions for outcomes conditioned on latent variables. In this paper, we present a more flexible alternative that allows the latent conditional distribution to be a mixed member of finite Gaussian mixtures. Our model was motivated by our efforts to describe common styles of return impact location of professional tennis players and is the reason we name the approach a `latent style allocation' model. In a fully Bayesian implementation, we apply the model to 142,803 return points played by 141 top players at Association of Tennis Professional events between 2018 and 2020 and show that the latent style allocation improves predictive performance over a finite Gaussian mixture model and identifies six unique impact styles on the first and second serve return. 
\end{abstract}


\section{Introduction}

The last two decades have witnessed an explosion in the use of tracking data systems in professional sports \cite{chase2020data}. Tracking systems capture fine-grained spatiotemporal information during competitive events and provide detailed summaries of player performance. Analysis of tracking data is regarded as a critical tool for gaining a competitive edge in sport and, as such, has become the central focus of an emerging sports analytics industry \cite{morgulev2018sports,gerrard2016analytics}.

Tracking systems began to be used at professional tennis events in the late 2000s for line call review \cite{fischetti2007or}. In the past decade, the use of camera-based tracking systems have become a mainstay of major events, resulting in the collection of detailed positional data for hundreds of men's and women's tennis matches each year. Despite the wealth of positional data in the sport, the proprietary restrictions on these data have been a hindrance to research into tennis performance. As a result, only a limited number of studies have described the spatiotemporal characteristics of shots in tennis \cite{kovalchik2018shot}, and most have focused only on the serve \cite{wei2015predicting,hizan2015gender,mecheri2016serve}.

The serve return is the first shot of the receiver, making it the most important receiving shot in tennis. Despite it's importance, few studies have considered the serve return characteristics of professional players. Hizan and colleagues looked at the distribution of landing location of elite junior players by categorizing the location into three different zones from which they observed that 50-70\% of serve returns were hit to the middle of the court \cite{hizan2014comparative}. Reid, Morgan and Whiteside examined individual summary statistics of the serve return among Australian Open players, including depth of position and contact height \cite{reid2016matchplay}. Earlier studies had spatial data of the serve return yet restricted their analyses to univariate summaries of the physical properties of the tennis shot. To our knowledge, no prior work has directly modeled the multi-dimensional spatial characteristics of the serve return.

Across sports, a number of strategies for modeling spatial data have emerged in recent years. Popular strategies involve coarsening of spatial coordinates\textemdash either as counts on a segmented field \cite{miller2014factorized,yue2014learning} or as image masks for applications in computer vision \cite{fernando2019memory,nistala2019using}\textemdash each losing information from the outset. Owing to the complexity of spatial data in sport, a number of authors have proposed non-parametric methods to model tracking data. Gaussian processes have been a common non-parametric method in sports applications, with multiple examples in basketball \cite{cervone2016multiresolution} and soccer \cite{durante2014bayesian,bojinov2016pressing}. The major strength of such non-parametric approaches is their flexibility. However, that flexibility often comes at the cost of interpretability and scalability \cite{liu2020gaussian}. 

Recently, mixture models have emerged as a more scalable and interpretable model-based framework for spatial data problems in sport. One of the most appealing features of mixture models for sports applications is that they directly embed a latent group factor, which can be interpreted as an unobserved subgroup category. Gaussian mixture models have been used in tennis to build a taxonomy of shots \cite{kovalchik2018shot} and a generative model for shot events \cite{kovalchik2020space}. Dutta, Yurko and Ventura (2020) applied finite Gaussian mixture models to discover coverage types for passing plays from NFL tracking data \cite{dutta2020unsupervised}. Hu, Yang and Xue (2020) used a log Gaussian Cox process with a mixture of finite mixtures to describe shooting styles among NBA players \cite{hu2020bayesian}. 

There is a growing recognition of the advantages of mixture models for sports spatial data. Yet current mixture model approaches have a major limitation: the conditional distribution for the spatial outcome, conditioned on the latent category, follows a standard multivariate distribution. As a result, latent clusters are modelled with parametric distributions which may be overly restrictive for some spatial applications. In this paper, we address this limitation by introducing a Gaussian mixture model with latent style allocation, where the style-specific distribution is itself a finite mixture of multivariate Gaussians. We apply this model to newly released public data on the return impact position of professional tennis players and show that it improves the unsupervised classification of the styles of player positioning when receiving serve.

\section{Tennis Terms}

A singles tennis match is a contest between two players played on opposite sides of a net on a court with an inbounds area that is 23.78 meters long and 8.23 meters wide. In professional tennis, the dimensions of the tennis court do not change but the court surface can be any one of three major surface types: clay, grass, or hard court.

Every tennis match consists of multiple sets and multiple games within a set. In each game, one player is the server and the other is the receiver, the roles alternating with each game. The serving player takes the first shot of the point which is called the `serve'. A serve is typically an overhead shot and it must land in the service box\textemdash a rectangle on the opposite side of the net of dimensions 6.4 meters deep and 4.115 meters wide\textemdash without touching the net for the point to proceed. Odd points are served to the service box located right of the center on the receiver's side (the `Deuce' side) and even points are played to the service box located left of center (the `Ad' side). A server has two attempts to put the serve in play. If the second serve attempt is not in play, the server has committed a `double fault' and automatically loses the point. A serve that is in play may or may not be returned by the receiver. The receiver is said to `return` the serve when they hit a shot off the serve that goes over the net and lands inbounds.

\section{Exploration of Return Impact Data}

The Association of Tennis Professionals (ATP) is the main organizer of tournaments in men's professional tennis. Beginning in 2018, the ATP began to provide summaries of tracking data on its website, including the location of ball at impact on the serve return (\texttt{www.atptour.com}). The data on returns include shots ending in an error or shots in play. When the receiver does not make contact with the serve, when a serve is an ace, for example, there is no return impact information. The data for the present study comprises return impact location, serve number, serving player, returning player, court side, match surface, event name and date for all matches with published tracking data between 2018 and 2020. Matches with 30 or more return points were considered complete and were retained for analysis. Further, to focus on top players, only receivers with 3 or more matches were included. After applying these inclusion criteria, the final sample included 141 receiving players, 1,334 matches, and 142,803 return points.

\begin{figure}
\centering
\includegraphics[width=0.5\textwidth]{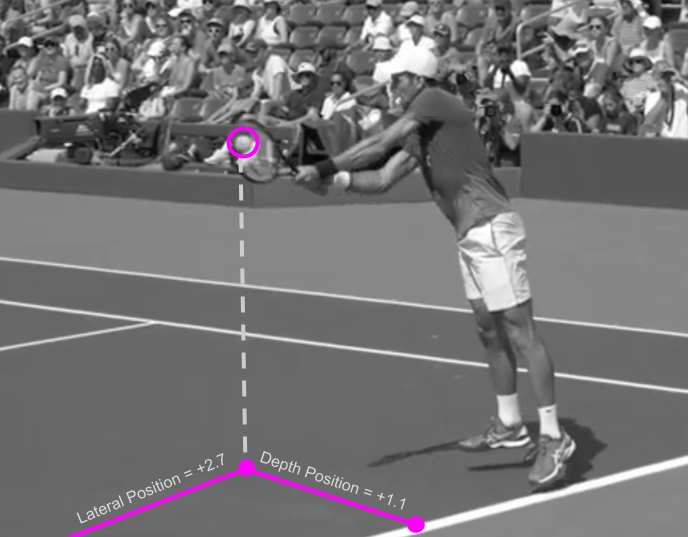}
\caption{Illustration of the return impact location and its 2D spatial measurement. The `lateral position` is the length in meters of a straight line from the ball location at impact to the centre line of the court. The `depth position` is the length in meters of a straight line from the ball location to the baseline, with positions beyond the baseline taking a negative value.} 
\label{fig:return_impact}
\end{figure}

The `return impact' is the event when the receiver's racquet makes contact with the ball on the serve return. The 2D position consists of the lateral and longitudinal location of the ball, which will be referred to as the `lateral position'  and `depth position', respectively. All coordinates are in meters. For the lateral position, the coordinate is the distance from the centre line down the middle of the court, with negative values being left of center and positive values right of center. The depth position is the distance from the baseline with positive values indicating a location inside the court and negative values a location beyond the baseline. An illustration of some of the return impact data is shown in Figure~\ref{fig:return_impact}.

\begin{figure}
\centering
\includegraphics[width=0.8\textwidth]{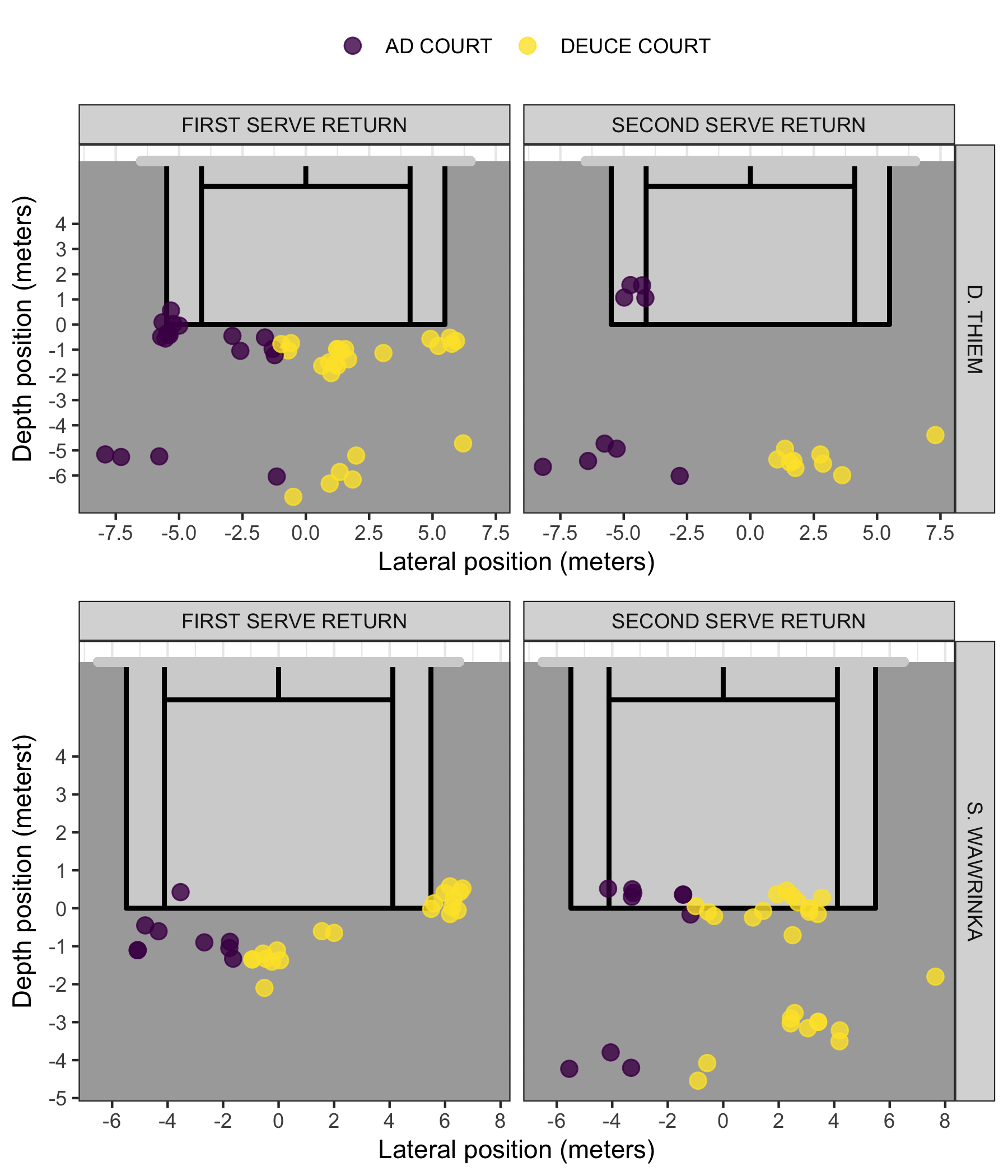}
\caption{Return impact locations from a single match for two illustrative players: Dominic Thiem and Stan Wawrinka.} 
\label{fig:player_sample}
\end{figure}

A typical sample of the return impact locations is shown in Figure~\ref{fig:player_sample}. The first row of plots show the return impact locations for Dominic Thiem for one clay court match against Roger Federer played in 2019. On both first serve (to the left) and second serve (to the right), Thiem's impact locations show strong evidence of clustering where the region within 2-3 meters behind the baseline was rarely the depth of impact in this match. Multiple modes in both the lateral and depth dimensions are also seen in the lower panel that shows sample positions of Stan Wawrinka from one hard court match played against Andy Murray in 2019. Wawrinka was more often inside the court on the second serve return than the first serve return, though he did make impact at depth of 3 meters or more behind the baseline on some second serves. 

\begin{figure}
\centering
\includegraphics[width=0.8\textwidth]{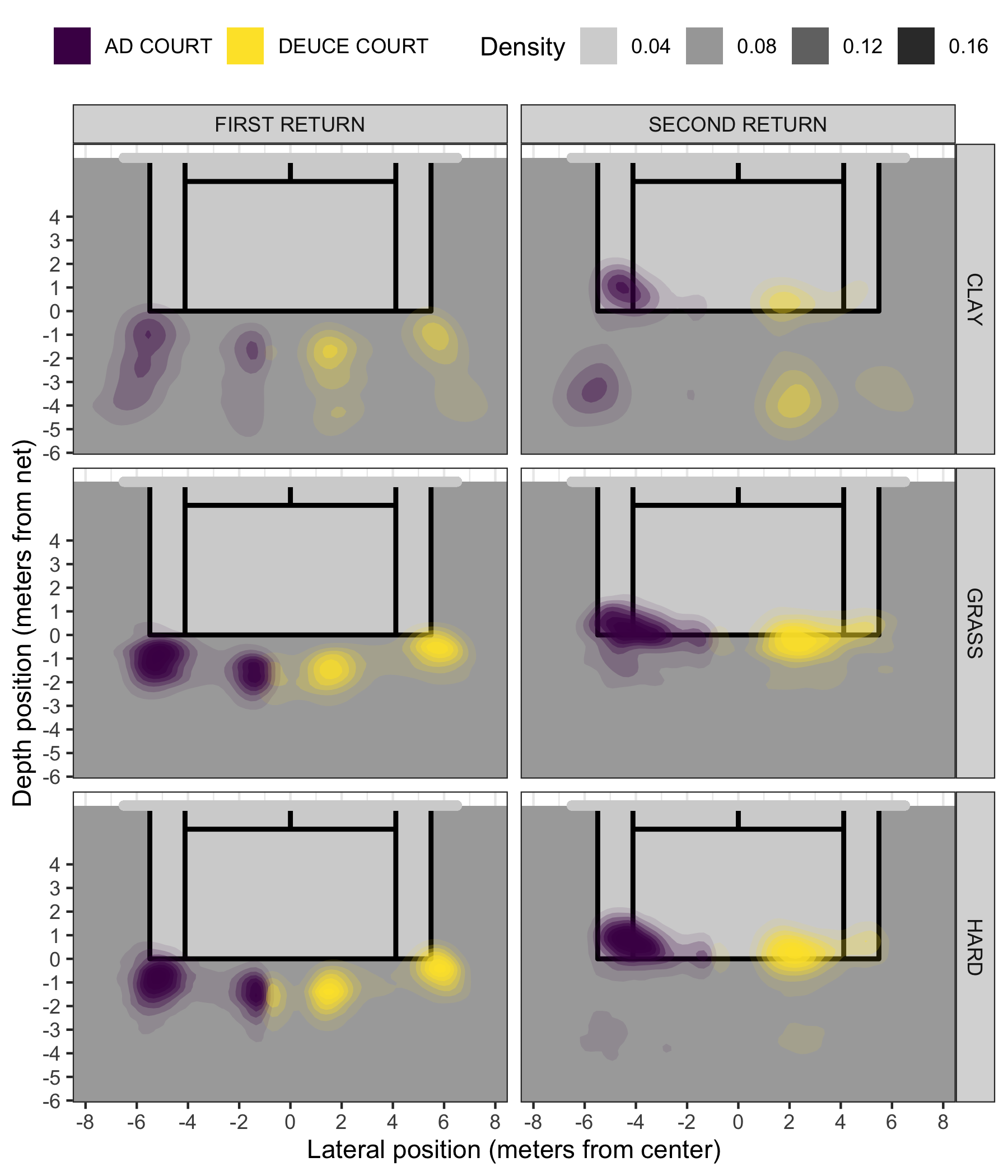}
\caption{2D contour density of all return impact locations in the study sample by serve number and court surface.} 
\label{fig:tour_contours}
\end{figure}

Figure~\ref{fig:tour_contours} summarizes the regions of highest density in 2D space over all of the return points in the study sample. On first serve return, we observe strong clustering in the lateral dimension, with the region of 2-4 meters from the center of the court infrequently the position of impact. There are marked differences in the depth of position on the first serve return by surface, where the highest probability of depths of more than 3 meters beyond the baseline occurs on clay courts. In general, impact locations on the second serve return show less clustering along the lateral dimension but more in the depth dimension, especially on the clay surface.

The empirical summaries presented in this section highlight several key properties of the spatial distribution of return impact. Both the modes and variances of return impact distribution appear to be influenced by surface, serve type and player. Even within player, it is possible to observe clustering in both the lateral and depth dimensions that is not explained by match-to-match variation or serve type and would not be well-described by a multivariate normal distribution.

\section{Model}

\subsection{Finite Mixture}

Let $\mathbf{Y}_{ij} = (x_{ij}, y_{ij})$ be the $i$th receiver's coordinate for the return position on their $j$th return impact point. One of the simplest models to describe the spatial distribution of $\mathbf{Y}_{ij}$ is a multivariate normal, $\mathbf{Y}_{ij} \sim MVN(\boldsymbol{\mu}, \boldsymbol{\Sigma})$. This model would make the strong assumption that all return positions are generated by a common mean and covariance. However, given the non-standard patterns of the return impact positions observed in our exploratory analysis, we seek a less restrictive descriptive model. 

One extension of the simple MVN model is to suppose that there are $K$ unknown return styles, each with their own mean and covariance, $(\boldsymbol{\mu}_k, \boldsymbol{\Sigma}_k)$. In this case, each observation has a style type, $k_{ij}$, and the generative model is conditional on the style type. Namely, 

$$
\mathbf{Y}_{ij} | k_{ij} = k \sim MVN(\boldsymbol{\mu}_k, \boldsymbol{\Sigma}_k).
$$

\noindent The style group is not actually known but is instead a latent quantity. For this reason, the style groups are given their own distribution. 

$$
k_{ij} \sim Categorical(\Theta_K)
$$

\noindent In this case, the total number of styles $K$ is treated as known and all $k_{ij}$ are generated from a common distribution. This is an example of a finite mixture model. 

A key assumption of the finite mixture model and popular extensions of it, including the mixed membership model, is that the group conditional distribution is still a standard density. For the present example, this equates to the assumption that player return styles can be well described by a multivariate normal. Yet our exploratory analysis shows that, even within individual player, there is clustering in return position that would not be adequately represented by any standard multivariate density. Thus, we seek an extension of standard mixture approaches that can account for within-style clustering in return impact position.

\subsection{Latent Style Allocation}

The latent style allocation model introduces a more flexible description of return impact styles by modeling styles as a mixture distribution. This is accomplished via a two-level latent variable mixture model. The highest latent variable is the style factor consisting of $K$ styles. Each style, in turn, has a unique probability vector of $M$ return impact patterns. This can be represented by the following conditional generative process,

\begin{align}
\boldsymbol{\theta}_k &\sim G(.) \\
\boldsymbol{\pi}_{i} &\sim Dirichlet(\boldsymbol{\alpha}_0) \\
k_{ij} |\boldsymbol{\pi}_i &\sim Categorical(\boldsymbol{\pi}_i) \\
m_{ij} |k_{ij} &\sim Categorical(\boldsymbol{\theta}_{k_{ij}}) \\
\mathbf{Y}_{ij} | m_{ij} &\sim MVN(\boldsymbol{\mu}_{m_{ij}}, \boldsymbol{\Sigma}_{m_{ij}})
\end{align}

\noindent  Patterns are draw conditional on the style category, $k$. Styles, by contrast, are determined by the receiver, each receiver having his own probability distribution over the $K$ possible styles, $\boldsymbol{\pi}_i$. In this way, the latent style allocation model can be viewed as an extension of the Gaussian mixed membership (GMM) model, as each receiver can be a member of a mix of styles. 

In contrast with the GMM, the latent style allocation treats playing styles as a mixture of patterns and player outcomes as a mixture of style types. A direct consequence of this key distinction is that players within the same style group share information, each style being defined by its pattern simplex. A player-specific marginal distribution is achieved by giving each player his own probability distribution over the style groups. 

Covariate effects are introduced with the specification of the mean parameters $\boldsymbol{\mu}_m$. Given a vector of covariates $\mathbf{x}_{ij}$, we model the mean as the sum of pattern-specific effects and receiver and server offsets. Specifically, 

\begin{equation}
\boldsymbol{\mu}_{m_{ij}} = (\boldsymbol{\alpha}_{m_{ij}} + \boldsymbol{\eta}_{r(ij)} - \boldsymbol{\delta}_{s(ij)}) \mathbf{x}_{ij}
\end{equation}

\noindent where $\boldsymbol{\alpha}_m$ is a $D \times P$ set of population effects, $\boldsymbol{\eta}_r$ is a matrix of equal dimension having receiver effects, and $\boldsymbol{\delta}_s$ a matrix of equal dimension with server effects. To illustrate the mean structure, suppose the $\mathbf{x}$ includes an intercept and indicator for a clay court match. The mean for a hard court match would then be represented as $\mathbf{x} = (1, 0)$ and, writing  $(\boldsymbol{\alpha}_m + \boldsymbol{\eta}_{r} - \boldsymbol{\delta}_{s}) = \begin{bmatrix} a_{11} & a_{12} \\ a_{21} & a_{22} \end{bmatrix}$, the MVN mean becomes, 

$$
\boldsymbol{\mu}  =  \begin{bmatrix} a_{11} & a_{12} \\ a_{21} & a_{22} \end{bmatrix} \times \begin{bmatrix} 1 \\ 0 \end{bmatrix} = \begin{bmatrix} a_{11} \\ a_{21} \end{bmatrix}
$$

\noindent Each of the $\boldsymbol{\alpha}_m$, $\boldsymbol{\eta}_{r}$,  and $\boldsymbol{\delta}_{s}$ are given a multivariate-normal prior with zero mean and covariance that has a non-informative LKJ Cholesky prior, a common choice for more efficient Bayesian hierarchical models \cite{lewandowski2009generating}. The remaining parameter of the pattern-conditional MVN model is the observation-level covariance, $\boldsymbol{\Sigma}_m$. For the observation-level covariance, we modify the standard LKJ Cholesky prior to allow for possibly heavy-tailed outcomes. This involves drawing the scaling factors from a Student T distribution with 1 degree of freedom, truncated to the positive real line.

\begin{figure}
\centering
\includegraphics[width=0.6\textwidth]{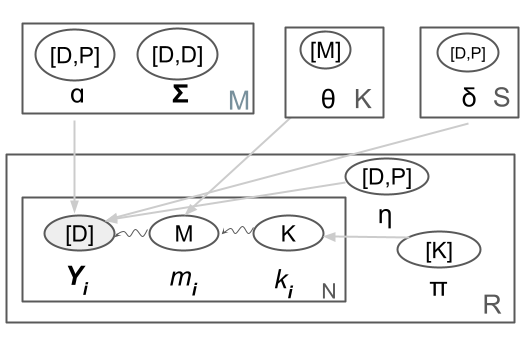}
\caption{Simplified plate diagram of latent style allocation model showing the generative distributions for the latent style factor $k$, latent pattern factor $m$ and the outcome model conditional on these latent factors.}
\label{fig:plate}
\end{figure}

The simplified plate diagram in Figure~\ref{fig:plate} illustrates the generative process for the latent style and pattern factors and how these determine the model for return impact locations. For each location, a style category $k$ is drawn according to the receiver-specific distribution over style types $\theta_r$. A pattern type is selected according to the style-specific probability distribution $\pi_k$. The pattern group determines the mean and covariance of the multivariate normal distribution with the mean effects adjusted according to receiver ($\eta$) and server ($\delta$) offsets. 

The process $G(.)$ is a parsimonious ordered stick-breaking procedure that ensures the identifiability of the latent style groups. The stick-breaking procedure begins with $K$ ordered standard normal random variables. An inverse-logit, denoted $g(.)$ below, transforms these into an ordered set of probabilities. A stick-breaking procedure is then sequentially applied as detailed below,

\begin{align}
\begin{split}
\boldsymbol{\beta}_{km} &\sim N(0, 1) \mbox{ and } \beta_{1m} << \beta_{2m} << \ldots << \beta_{Km}\; \mbox{for}\; m=1,\ldots,M-1 \\
\nu_{km} &= \mbox{logit}^{-1}(\boldsymbol{\beta}_{km}) \\
 \boldsymbol{\theta}_{km} &= \nu_{km} \prod_{l=1}^{m-1} (1 - \nu_{kl})
\label{eq:latent_mmm_stick}
\end{split}
\end{align}

\noindent With this process, a total of $K(M-1)$ parameters are used to define the pattern simplexes across the style groups. With the ordering constraints, the prior will encourage the first style group ($k=1$) to have less weight on the first pattern mixture while the final style group ($k=K$) will tend to put greater weight on the first pattern. In the simplest case of a two-component Gaussian mixture $(M=2)$, the first component mixture weight would increase monotonically from the first to last latent style group. Neither the ordering of the style groups nor the ordering of the pattern groups are directly influenced by the mean or covariance parameters of the observation-level multivariate normal.

The latent group modeling presented here has close parallels to topic models. In fact, the GMM is a direct analog of latent Dirichlet allocation for a continuous multivariate outcome. A flexible extension of the standard LDA adds a hierarchical Dirichlet process (HDP) to the generating distribution of topics, in which topics are a mixture of an unknown set of categories \cite{teh2006hierarchical}. This results in partial pooling between the known groups (i.e. documents) sharing the same category. When latent parameters follow a Dirichlet process, the number of latent groups is non-parametric. While this provides appealing flexibility, the countably-infinite set of latent categories presents difficulties for implementation in modern probabilistic programming languages, like Stan, where sampling of discrete parameters is not supported.  

Like the HDP, the latent style allocation model allows information sharing between the known groups. A key advantage of the latent style allocation model, however, is that is can be readily implemented in modern Bayesian programming languages. The reason for this stems from the form of the marginal likelihood. Focusing just on the likelihood contribution to the target posterior and letting $L_{ij}(\boldsymbol{\Theta})$ be the term for one return impact location given model parameters $\boldsymbol{\Theta}$, the marginal likelihood for a single data point is the following sum

\begin{equation}
L_{ij}(\boldsymbol{\Theta}) = \sum_{k=1}^{K} \sum_{m=1}^{M} \pi_{ik} \theta_{km} MVN(\mathbf{Y}_{ij}; \boldsymbol{\mu}_m, \boldsymbol{\Sigma}_m)
\end{equation}

\noindent where $\pi_{ik}$ is the $i$th receiver's probability weight for the $k$th style style and $\theta_{km}$ the probability weight for the $m$th pattern type within the $k$th style category. When $K$ and $M$ are fixed, it is possible to sum over the discrete variables corresponding to the latent style and pattern groups. There is a loss in flexibility when the number of latent groups are fixed but the trade-off, in this case, is a more computationally tractable posterior. 

With two levels of latent grouping, the latent style allocation model can introduce greater complexity by increasing the number of style components, the number of pattern components, or both. While the choice of components at either level can be guided by standard model selection practices for Bayesian inference, some intuition about the comparative impact of increasing the number of style versus the number of pattern components will be a useful guide for selection procedures. We note that the latent style group describes the behavior of a collection of players while the latent pattern group describes within-player behavior. Thus, a greater number of patterns would be appropriate when there is considerable within player heterogeneity in return impact locations, while a greater number of styles would be appropriate when there are greater between-player differences in return impact locations.

\section{Application}

Using the ATP data sample, we investigate the choice of style and pattern components in the latent style allocation model that provide the best trade-off between predictive performance and model complexity. This is done by evaluating the expected log pointwise predictive density for all combinations of style and pattern components from 2 to 8. ELPD is a Bayesian leave-one-out cross-validation measure that assesses how well a model is able to describe future observations \cite{vehtari2017practical}. Between two models fit to the same data, the model with ELPD statistically closer to zero should be preferred. The ELPD can be efficiently approximated with Pareto-smoothed importance sampling \cite{vehtari2016bayesian}.

Each latent style allocation model includes covariate effects for three serve directions on each court side and surface indicators for each of the three major surface types. First and second return points are fit separately, owing to the known heterogeneity between these two point types. In addition to identifiers for the receiver and serving player, these were the only contextual variables available from the source data. 

The performance of the selected latent style allocation model is benchmarked against three other related approaches: a multivariate normal (MVN), a finite mixture, and mixed membership model. The MVN model has no latent group variables and all $Y_{ij}$ are generated from an MVN with the same population effects and covariance. The finite mixture model is a simplification of the Gaussian mixed membership model where there is a constant pattern distribution for all players, $\boldsymbol{\pi}_i = \boldsymbol{\pi}$. In contrast to the finite mixture, the Gaussian mixed membership model draws a receiver-specific pattern distribution using a Dirichlet prior with no pooling of information between players. For these benchmarks, the number of mixture components in these alternative models was equal to the number of pattern components in the selected latent style allocation model. 

All models were fit using the Stan language \cite{carpenter2017stan} and its variational inference algorithm \cite{kucukelbir2015automatic}.

\section{Results}

Over all possible combinations of style groups and mixture components, the first serve and second serve return impact models with optimal ELPD had 6 style groups and 6 mixture components. Table~\ref{tab:elpd} benchmarks the ELPD of the selected latent style allocation models against the multivariate normal, finite Gaussian mixture and Gaussian mixed membership alternatives, where the number of mixture components was set to 6 across all mixture models. The latent style allocation model improved predictive performance over both mixture model alternatives. For first serve returns, we observed a 2.8\% improvement in predictive performance over a mixed membership model with the same number of mixture components; whereas a 3.5\% improvement was observed for second serve returns.

\begin{table} \centering
\caption{Comparison of model expected log pointwise predictive density.}
{\begin{tabular}{lcc} \toprule
 Model & First Serve Return & Second Serve Return \\ \midrule
 Multivariate Normal & -181748 & -146143 \\
 Finite Mixture  $(M=6)$ & -157105 &  -124560\\
 Mixed Membership $(M=6)$ & -146365 & -123107 \\
 Latent Style Allocation $(M=6,K=6)$ & -142358 & -118955 \\ \bottomrule
\end{tabular}}
\label{tab:elpd}
\end{table}

\begin{figure}
\centering
\subfloat[First serve return.]{%
\resizebox*{0.8\textwidth}{!}{\includegraphics{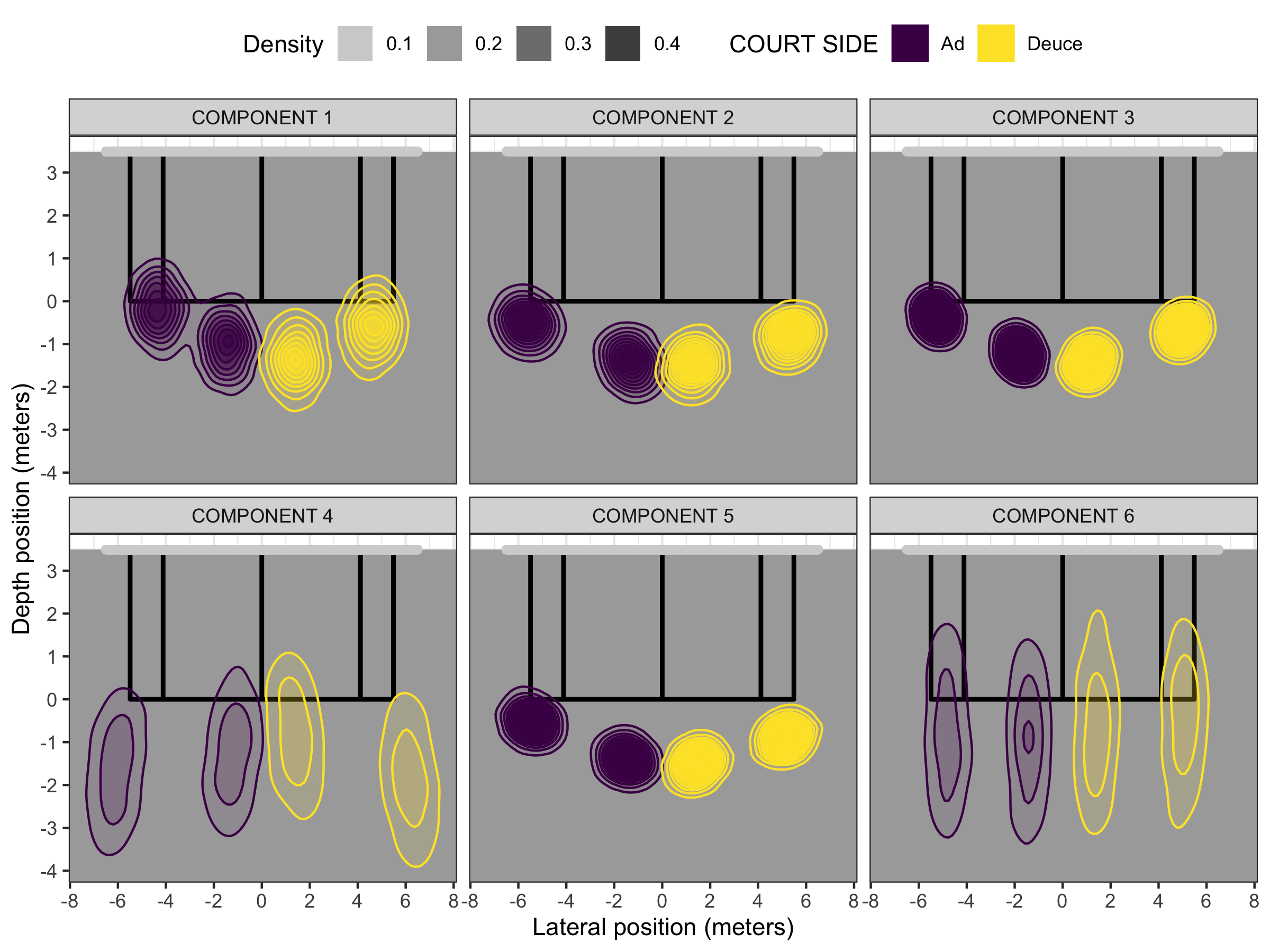}}}\hspace{5pt}
\subfloat[Second serve return.]{%
\resizebox*{0.8\textwidth}{!}{\includegraphics{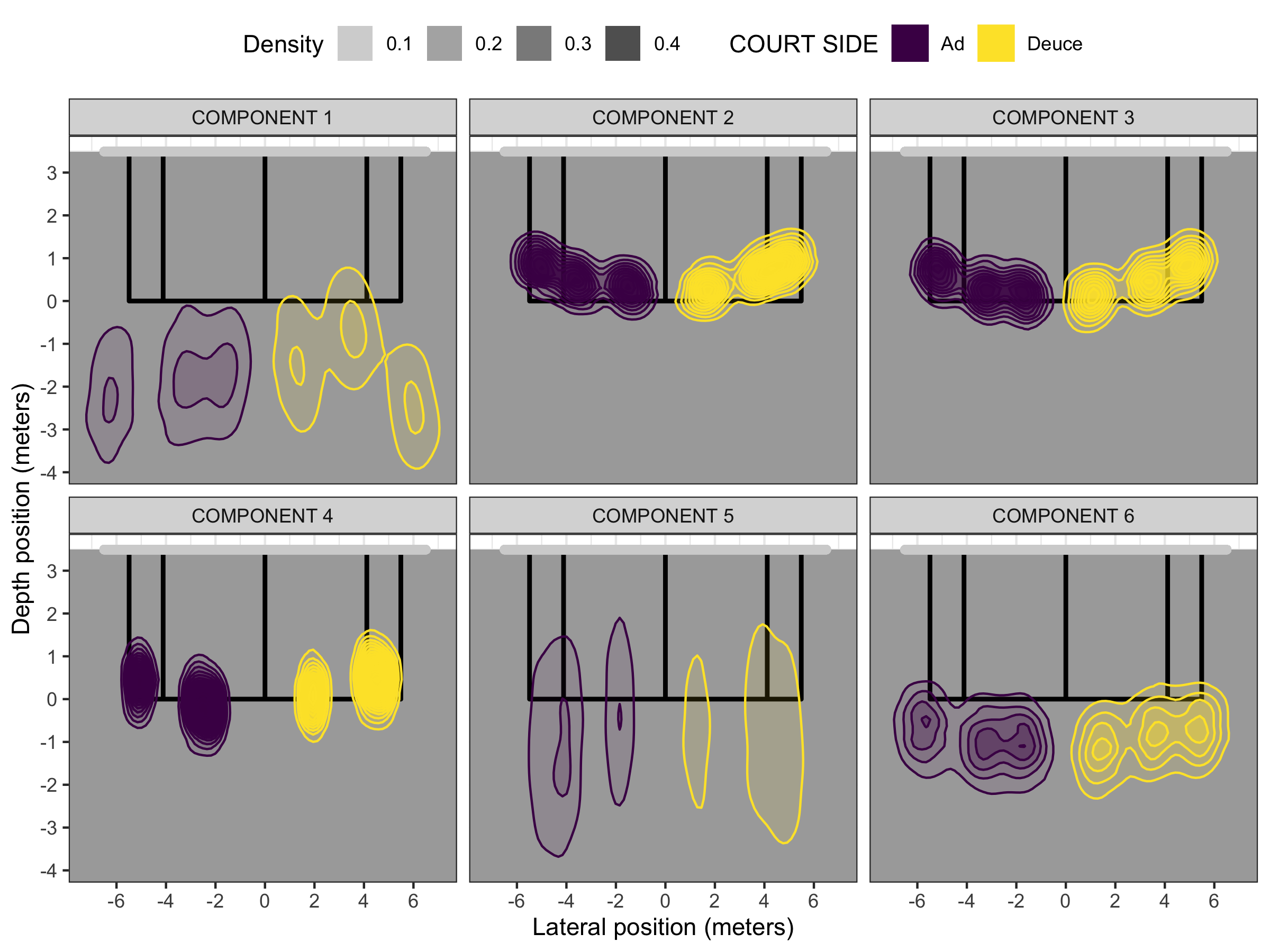}}}
\caption{Tour-level predictive posterior for hard court return impact locations for each mixture component of the latent style allocation model.} \label{fig:posterior}
\end{figure}

The tour-level posterior summaries represent the spatial distribution of return impact marginalized over all player effects. In this way, the tour-level posterior can be interpreted as the predicted impact locations for an average receiver against an average server. Figure~\ref{fig:posterior}~(a) gives the component-specific tour-level posterior distribution for first serve returns as a 2d contour plot. All of the distributions show two distinct clusters on each court side, which represent the impact of serves out wide and down the T (along the centre line of the court), respectively. There are three components with the lowest dispersion around each mode (components 2, 3 and 5), which have the greatest similarity among the component-specific densities. The most probable depths for these components are in the range of 0 to 1 meter beyond the baseline on wide serves and 1 to 2 meters for serves down the T. The region between 2 to 4 meters out wide is not frequently occupied for any of these components, but they do differ in the likelihood of the position at the center of the court, component 2 having the most concentration in the center compared to components 3 and 5.

Component 1 shares similar modes with component 5 but shows more dispersion along the depth dimension and towards the center of the court (Figure~\ref{fig:posterior}~(a)). Components 4 and 6 stand out for having the greatest dispersion among the mixture components in the depth dimension, where probable depths are in the range of 0 to 4 meters for component 4, and 1 meters inside the baseline to 3 meters beyond the baseline for component 6. We also observe that positions outside of the sidelines are common for component 4 but rare for component 6.

Using similar graphical summaries for the second serve return tour-level posteriors, we find more modes along the lateral dimension of the court, with returns of centrally positioned serves (`body' serves) being more common (Figure~\ref{fig:posterior}~(b)). Two components, components 2 and 3, have very similar distributions on hard courts, each with probable depths that are 1 to 0 meters inside the baseline and similar high-density regions along the width of the court at 1, 3 and 5 meters from the center. We note that there is a marked difference in depth on clay court for these component densities, and that is the main difference between them. 

Component 4 is the other low-dispersion component mixture (Figure~\ref{fig:posterior}~(b)). This mixture is distinctive for having probable depth of impact in the range of 1 meter inside to 1 meter behind the baseline. The most probable lateral positions are at 2 meters and 5 meters from the center. Component 5 has similar lateral modes but higher variance in depth, with depth locations ranging from 1 meter inside to 4 meters behind the baseline. Component 6 is unique for having a concentration of depth of position in the range of 0 to 2 meters behind the baseline across the possible lateral positions. Finally, component 1 stands out in having high dispersion in both dimensions and covering the most range for impact locations beyond the baseline and beyond the sidelines of the court.

\begin{figure}
\centering
\subfloat[First serve return.]{%
\resizebox*{0.8\textwidth}{!}{\includegraphics{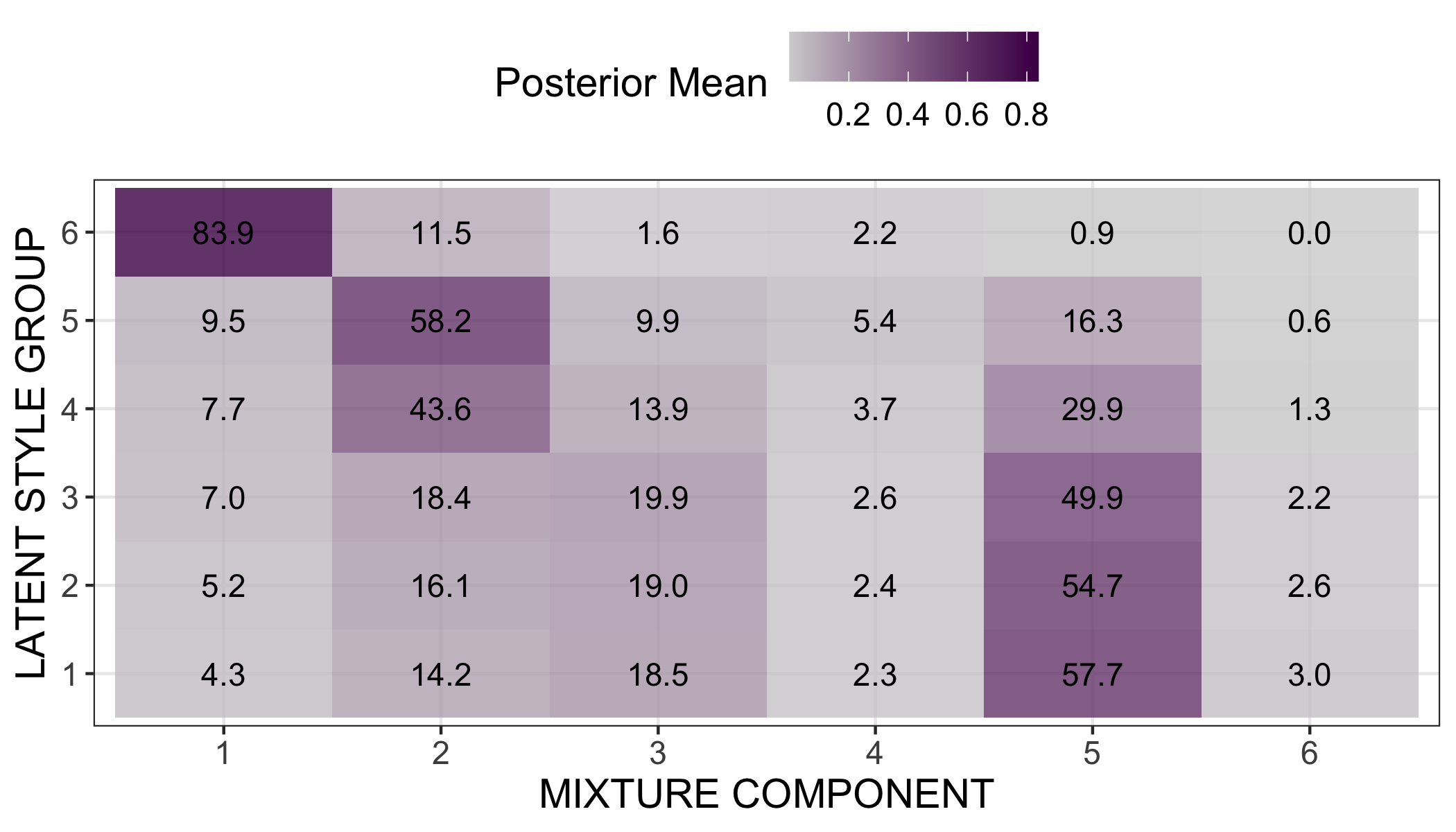}}}\hspace{5pt}
\subfloat[Second serve return.]{%
\resizebox*{0.8\textwidth}{!}{\includegraphics{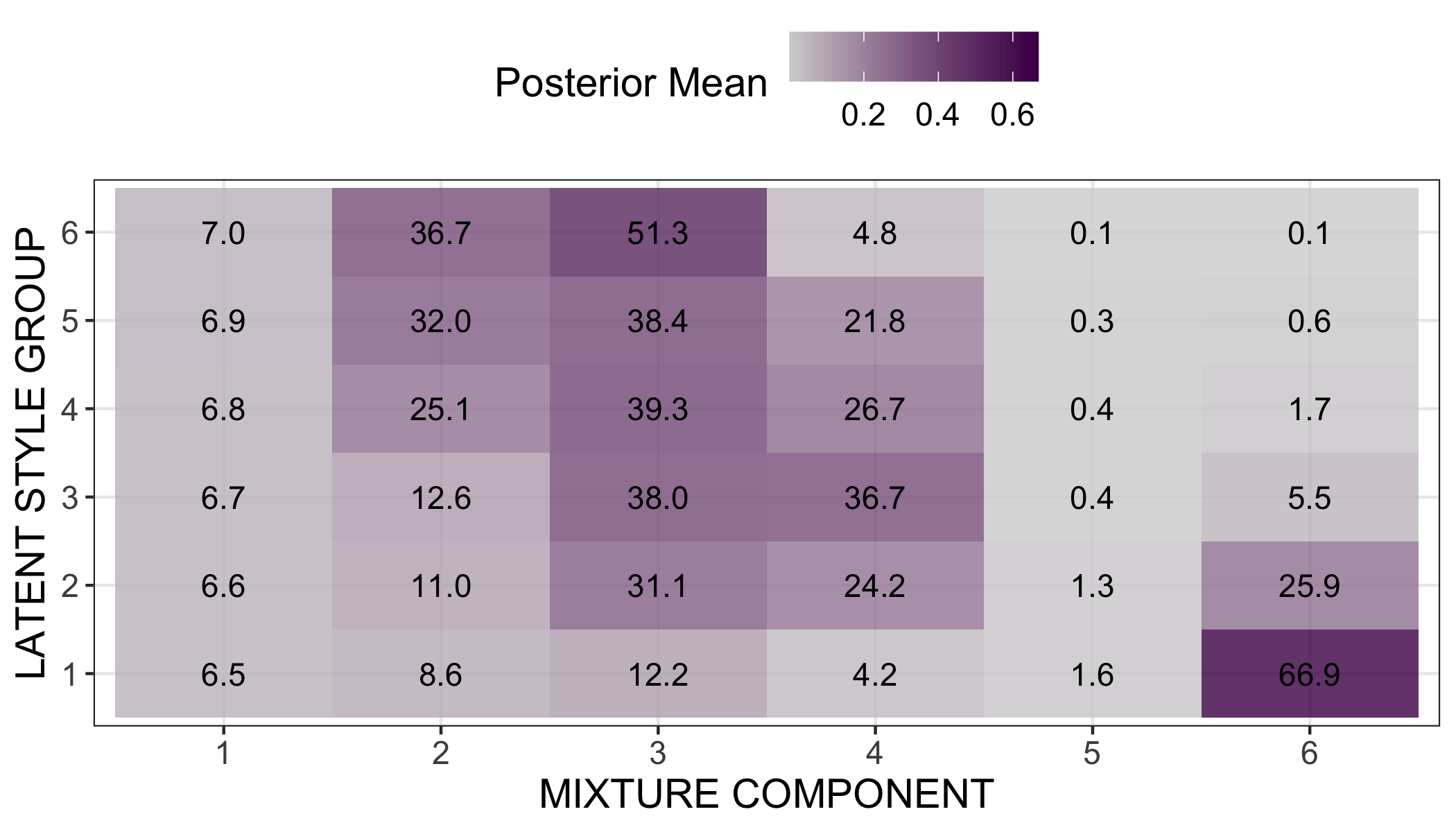}}}
\caption{Posterior mean of mixture component weights by latent style group.} \label{fig:style_mixtures}
\end{figure}

Differences among the style types is best summarized by comparisons of their mixture probability distributions. Figure~\ref{fig:style_mixtures} (a) shows the posterior mean weight for each mixture and style group for first serve returns. Style groups 1 to 3 have the plurality of weight on component 5, with components 2 and 3 taking the majority of the remaining weight. Style group 1 places the greatest weight on component 5 (58\%), whereas style group 3 puts the least (50\%). Styles 4 and 5 put the plurality of mixture weight on component 2 (44\% vs 58\%) and the second-highest on component 5 (14\% vs 10\%). Style 6 places 84\% weight on component 1, the highest weight of any style category, with the majority of the remaining weight going to component 2.

For second serve returns, style group 1 has the highest weight on any single mixture, which is allocated to component 6. Style groups 2, 4 and 5 each place 25 to 30\% weight on three components. Style groups 3 and 6 allocate the majority of weight to two mixtures, for style type 3, 74\% is split between components 3 and 4; whereas style type 6 allocates 87\% of its mixture density to components 2 and 3.

\begin{table} \centering
\caption{Summary of style group with maximum probability weight, n players (\% of 141 players) }
{\begin{tabular}{l cccccc} \toprule
  Style Group & 1 & 2 & 3 & 4 & 5 & 6 \\ \midrule
 First Serve Return  &  80 (56.7\%) &  17 (12.0\%) &  7 (4.9\%) & 4 (2.8\%) &  8 (5.7\%) & 25 (17.7\%) \\
 Second Serve Return  & 72 (51.0\%) & 11 (7.8\%) & 14 (9.9\%) & 10 (7.1\%) & 11 (7.8\%) & 23 (16.3\%) \\ \bottomrule
\end{tabular}}
\label{tab:max_style}
\end{table}

Summaries of the player-specific distribution of style weights suggest that style group 1 contributed the highest weight for the largest number of players on first and second return (Table~\ref{tab:max_style}). On first serve return, 80 of 141 players had the majority of their posterior style weight in group 1; on second serve return, 72 of 141 players had the majority of their style weight allocated to group 1. The second style category to have the plurality of weight was style group 6, again, for both the first and second return. The remaining style groups represented the plurality of the style allocation for 3-12\% of players.

\begin{figure}
\centering
\subfloat[First serve return.]{%
\resizebox*{0.8\textwidth}{!}{\includegraphics{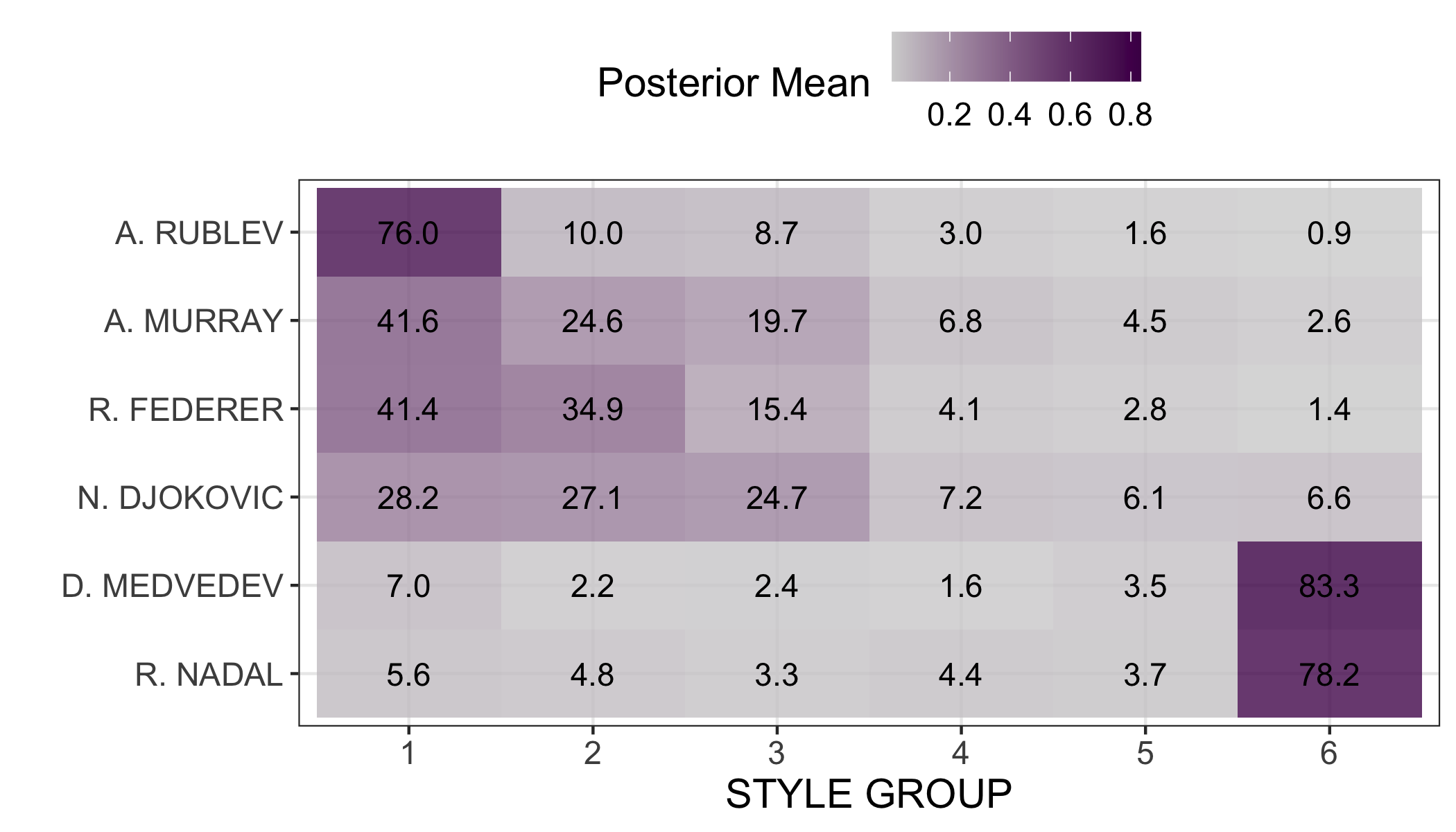}}}\hspace{5pt}
\subfloat[Second serve return.]{%
\resizebox*{0.8\textwidth}{!}{\includegraphics{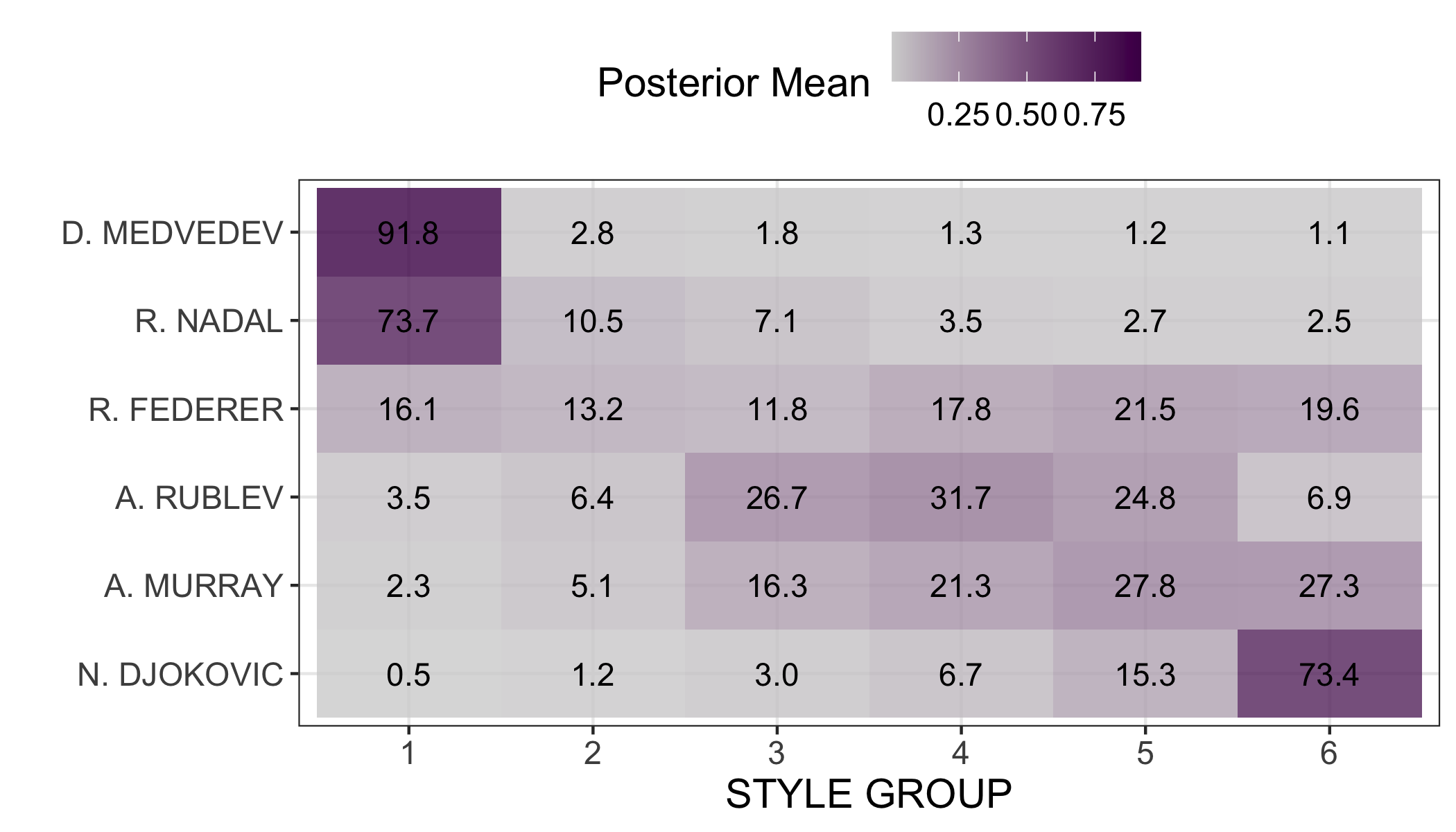}}}
\caption{Posterior mean of latent style probability distribution for selected top male players.} \label{fig:player_styles}
\end{figure}

\begin{figure}
\centering
\subfloat[First serve return.]{%
\resizebox*{0.6\textwidth}{!}{\includegraphics{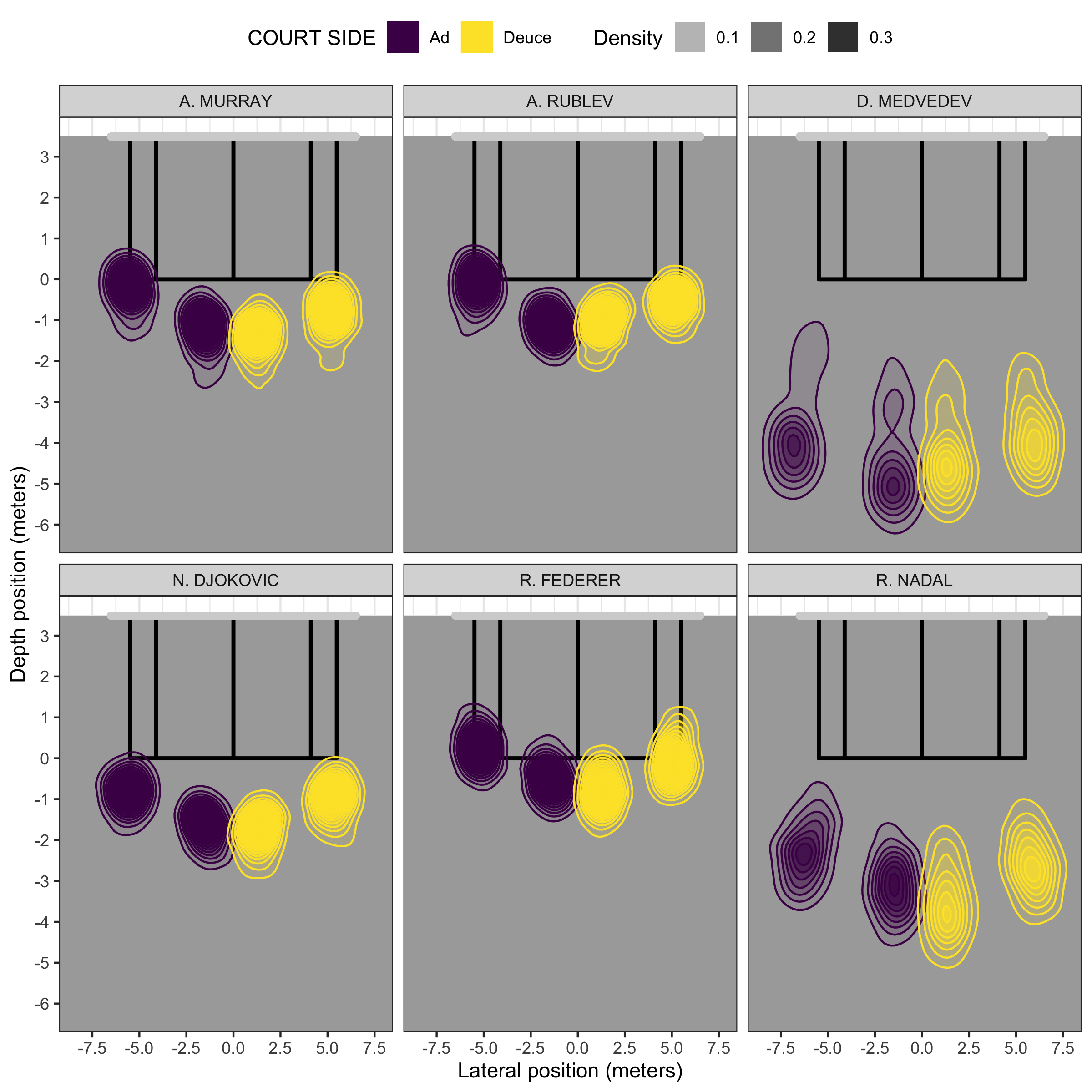}}}\hspace{5pt}
\subfloat[Second serve return.]{%
\resizebox*{0.6\textwidth}{!}{\includegraphics{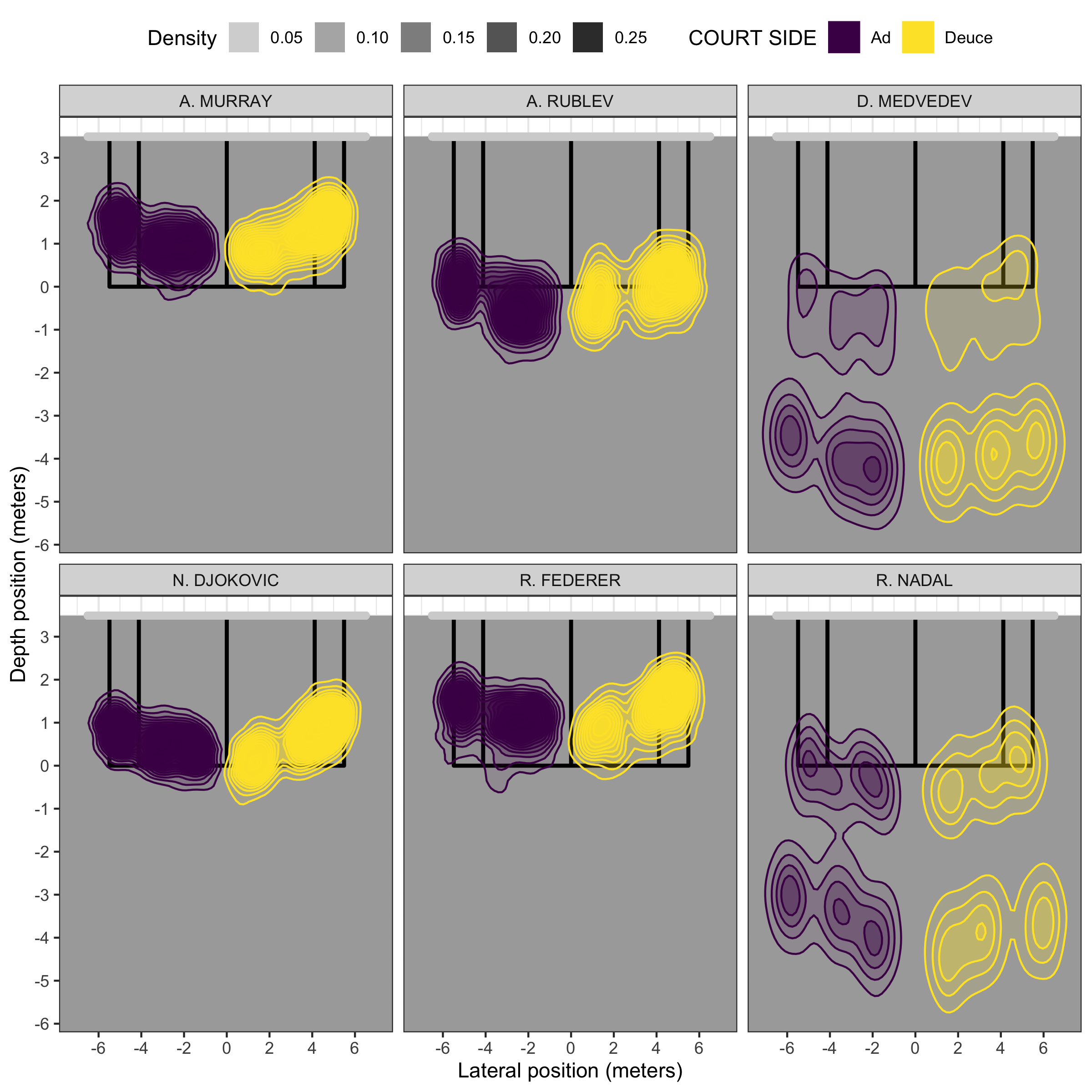}}}
\caption{Predictive posterior distributions of return impact locations on hard court for selected top male players.} \label{fig:player_contours}
\end{figure}

To illustrate how the latent style allocation model can assist with player-specific evaluation, we examine the posterior distributions for a selected group of top male players including 4 of the most successful players in tennis history: Andy Murray, Novak Djokovic, Roger Federer and Rafael Nadal. Figure~\ref{fig:player_styles} summarizes the posterior mean weights for each player and style group. Nadal and Medvedev have the most similar style distributions, with 78-83\% weight allocated to style group 6 on first serve and 74-92\% weight allocated to the first style group on second serve return. From the predictive posteriors on hard court shown in Figure~\ref{fig:player_contours}, we see that these styles stand out as having some of the deepest locations and most spread along the depth dimension on first serve (a), and two distinct modes\textemdash one at 3-4 meters behind the baseline and the other 0 to 1 meter behind the baseline\textemdash on second serve return. 

The remaining four players stand out in unique ways in their return impact patterns. Novak Djokovic's style on first serve is a mixture of three primary style groups, which amount to a depth of position that is between the aggressive locations of Roger Federer and the more defensive locations of Rafael Nadal (Figure~\ref{fig:player_styles} (a)). On second return, by contrast, Djokovic's patterns are best represented by style group 6, where the most frequent locations at the center of the court are at the baseline and 1 meter inside the baseline when further out wide (Figure~\ref{fig:player_styles} (b)). 

Federer and and Djokovic share similar characteristics in the lateral dimension of their first serve return, both tending to be further inside the court on wide serve and showing a high degree of symmetry by court side. Federer, however, is 0.5 to 1 meter further inside the court overall compared to Djokovic, making his return impacts the most aggressive among these players. That pattern is similar on second return, where the probability that Federer makes a return impact beyond the baseline is highly unlikely.

Murray and Rublev have a plurality of weight on style group 1 on first serve return (Figure~\ref{fig:player_styles} (a)). From their predictive posteriors, this allocation results in depths of 1-2 meters behind the baseline in the center of the court but shallower positions for serves out wide. These players stand out for having a clear asymmetry in their impact patterns by court side, where they tend to be further inside and outside the singles sideline on wide serves to the Ad side than to the Deuce side (Figure~\ref{fig:player_contours} (a)). Despite similar first serve return patterns, Murray and Rublev show stark differences in their second return styles. Murray's style weights are split fairly uniformly across 4 style categories and his predictive posterior shows impact locations that are 1 to 2 meters inside the court (Figure~\ref{fig:player_contours} (b)). Rublev has the majority of his style weights allocated to three groups and his predictive posterior shows a more conservative depth of impact, especially for centrally located serves to the Ad side.

\section{Discussion}

This paper has introduced a Gaussian mixture model with latent style allocation that provides a strategy for grouping multidimensional Gaussian mixtures across a finite number of latent groups. The model was motivated by our effort to describe patterns in the spatial distribution of return impact locations in men's professional tennis. In applying this model to thousands of points of top male players, we demonstrated that player return impact styles are better described by a mixture distribution than standard parametric distributions.  We also showed that the latent style factor provides a highly interpretable method for identifying and summarising more or less similar spatial patterns among players.

It is well known that players have distinct patterns when receiving serve. Jimmy Connors and Andre Agassi, for example, were both considered `aggressive returners' because they would take more shallow positions and make impact with the ball earlier than many of their contemporaries \cite{rutherford2017skills}. Although it is generally acknowledged that some top players prefer a deeper position on the serve return, detailed study of these preferences has been limited. To our knowledge, this is the first examination of return impact location of top tennis players in 2D space. Our application confirms that depth is an important property of serve return patterns, but it is not sufficient to describe the different impact styles of elite players. On first serve return, we find styles that overlap along the depth dimension but differ in their lateral positions. This study also identified styles with clear asymmetries in the lateral position on Ad and Deuce court that we have not seen discussed in the tennis coaching literature. We also observe that styles with deeper positions tend to be more diffuse, in general, suggesting that there is more range in the positioning and movement along the length of the court among top players. 

On second serve returns, players are often encouraged to take a more aggressive position \cite[p. 76]{antoun2007women}. While we observe a general shift forward in depth on second returns, this tendency is not universal. We find aggressive and defensive styles on both the first and second return that suggest that any combination among these is possible. A more general distinction we found in second return styles is the presence of distributions with multiple depth modes, with modes separated by several meters. Deeper exploration of the match data for some of the players in these style categories revealed that this kind of switching in position can happen within the same match, seemingly haphazardly. Some instructional texts recommend that players use different forward and backward positions on the serve return to `upset your opponent’s rhythm' \cite[p. 299]{smith2017absolute}, which could explain the pattern we observed. This is one of the more surprising results of the present study and worthy of further investigation. 

Our in-depth summary of a selected group of some of the most successful male tennis players in the past decade showed a wide range of return styles among the best in the game. Whether this is evidence that top players can be highly effective with any one of a number of return styles or that there is still room for even the best players to improve their return tactics remains an open question.

A key finding in this paper was the improved performance of the latent style allocation model over finite mixture alternatives. This result suggests two main conclusions. First, we observed multimodality in the lateral and depth dimensions that could not be explained by serve type or any other contextual variable available to us. This was most dramatic on second serve returns where a subset of players are observed to have high-density regions in the depth dimension that were separated by several meters. In this case, the latent style spatial descriptions required a more flexible distribution than what a standard parametric density could provide. Secondly, even among the top 150 male players, there is a wide range in the sample sizes of tracking data, as tracking systems are only available at the best events and marquee courts where lower-ranked players are less often observed. With sparse observations for some players, partial pooling of player distributions through the latent style groups can yield better predictive performance than a mixed membership model with no pooling in the assignment of player mixture weights.

More flexible frameworks for mixture models could also provide similar performance gains in modeling return impact locations over parametric or finite mixture alternatives. An HDP would be one of the most flexible options for a latent group factor variable that would allow for partial pooling and an unspecified number of group categories. The main advantage of the latent style allocation over the HDP is its ease of implementation. With modern Bayesian probabilistic programming languages, like Stan, sampling of an infinite process is often not possible without approximations \cite{nishimura2017discontinuous}. 

While the spatial distribution of return impact locations in tennis was the primary motivation of the latent style allocation model, we believe this model can be useful for more general spatial problems in sport. As the capture of spatiotemporal data in sport continues to grow, there will be a growing interest in models that can describe complex patterns of athlete movement in space and categorize styles of movement in a model-based way. The flexibility, interpretability, and accessibility of the latent style allocation model all make it a practically meaningful tool for this emerging area of performance analysis in sport.

\section*{Disclosure statement}

The authors have no conflicts of interest to disclose.

\bibliographystyle{acm}

\begin{thebibliography}{10}

\bibitem{antoun2007women}
{\sc Antoun, R.}
\newblock {\em Women's Tennis Tactics}.
\newblock Human Kinetics, Champaign, Illinois, 2007.

\bibitem{bojinov2016pressing}
{\sc Bojinov, I., and Bornn, L.}
\newblock {\em The pressing game: Optimal defensive disruption in soccer}.
\newblock 10th MIT Sloan Sports Analytics Conference, Boston, MA, 2016.

\bibitem{carpenter2017stan}
{\sc Carpenter, B., Gelman, A., Hoffman, M.~D., Lee, D., Goodrich, B.,
  Betancourt, M., Brubaker, M., Guo, J., Li, P., and Riddell, A.}
\newblock {Stan: A probabilistic programming language}.
\newblock {\em Journal of Statistical Software 76}, 1 (2017), 1--32.

\bibitem{cervone2016multiresolution}
{\sc Cervone, D., D’Amour, A., Bornn, L., and Goldsberry, K.}
\newblock A multiresolution stochastic process model for predicting basketball
  possession outcomes.
\newblock {\em Journal of the American Statistical Association 111}, 514
  (2016), 585--599.

\bibitem{chase2020data}
{\sc Chase, C.}
\newblock The data revolution: Cloud computing, artificial intelligence, and
  machine learning in the future of sports.
\newblock In {\em {21st Century Sports}}. Springer, Cham, Switzerland, 2020,
  pp.~175--189.

\bibitem{durante2014bayesian}
{\sc Durante, D., and Dunson, D.}
\newblock Bayesian logistic {Gaussian} process models for dynamic networks.
\newblock In {\em Artificial Intelligence and Statistics\/} (2014), PMLR,
  pp.~194--201.

\bibitem{dutta2020unsupervised}
{\sc Dutta, R., Yurko, R., and Ventura, S.~L.}
\newblock Unsupervised methods for identifying pass coverage among defensive
  backs with nfl player tracking data.
\newblock {\em Journal of Quantitative Analysis in Sports 16}, 2 (2020),
  143--161.

\bibitem{fernando2019memory}
{\sc Fernando, T., Denman, S., Sridharan, S., and Fookes, C.}
\newblock Memory augmented deep generative models for forecasting the next shot
  location in tennis.
\newblock {\em IEEE Transactions on Knowledge and Data Engineering 32}, 9
  (2019), 1785--1797.

\bibitem{fischetti2007or}
{\sc Fischetti, M.}
\newblock In or out?
\newblock {\em Scientific American 297}, 1 (2007), 96--97.

\bibitem{gerrard2016analytics}
{\sc Gerrard, B.}
\newblock Analytics, technology and high performance sport.
\newblock {\em Critical Issues in Global Sport Management 205\/} (2016),
  227--240.

\bibitem{hizan2015gender}
{\sc Hizan, H., Whipp, P., and Reid, M.}
\newblock Gender differences in the spatial distributions of the tennis serve.
\newblock {\em International Journal of Sports Science \& Coaching 10}, 1
  (2015), 87--96.

\bibitem{hizan2014comparative}
{\sc Hizan, H., Whipp, P., Reid, M., and Wheat, J.}
\newblock A comparative analysis of the spatial distributions of the serve
  return.
\newblock {\em International Journal of Performance Analysis in Sport 14}, 3
  (2014), 884--893.

\bibitem{hu2020bayesian}
{\sc Hu, G., Yang, H.-C., and Xue, Y.}
\newblock Bayesian group learning for shot selection of professional basketball
  players.
\newblock {\em {STAT}\/} (2020), e324.

\bibitem{kovalchik2020space}
{\sc Kovalchik, S., Ingram, M., Weeratunga, K., and Goncu, C.}
\newblock {Space-time VON CRAMM:} evaluating decision-making in tennis with
  variational generation of complete resolution arcs via mixture modeling.
\newblock {\em arXiv preprint arXiv:2005.12853\/} (2020).

\bibitem{kovalchik2018shot}
{\sc Kovalchik, S., and Reid, M.}
\newblock A shot taxonomy in the era of tracking data in professional tennis.
\newblock {\em {Journal of Sports Sciences} 36}, 18 (2018), 2096--2104.

\bibitem{kucukelbir2015automatic}
{\sc Kucukelbir, A., Ranganath, R., Gelman, A., and Blei, D.~M.}
\newblock Automatic variational inference in stan.
\newblock {\em Advances in Neural Information Processing Systems 2015\/}
  (2015), 568--576.

\bibitem{lewandowski2009generating}
{\sc Lewandowski, D., Kurowicka, D., and Joe, H.}
\newblock Generating random correlation matrices based on vines and extended
  onion method.
\newblock {\em {Journal of Multivariate Analysis} 100}, 9 (2009), 1989--2001.

\bibitem{liu2020gaussian}
{\sc Liu, H., Ong, Y.-S., Shen, X., and Cai, J.}
\newblock When {Gaussian} process meets big data: A review of scalable gps.
\newblock {\em {IEEE Transactions on Neural Networks and Learning Systems} 31},
  11 (2020), 4405--4423.

\bibitem{mecheri2016serve}
{\sc Mecheri, S., Rioult, F., Mantel, B., Kauffmann, F., and Benguigui, N.}
\newblock The serve impact in tennis: First large-scale study of big hawk-eye
  data.
\newblock {\em Statistical Analysis and Data Mining: The ASA Data Science
  Journal 9}, 5 (2016), 310--325.

\bibitem{miller2014factorized}
{\sc Miller, A., Bornn, L., Adams, R., and Goldsberry, K.}
\newblock Factorized point process intensities: A spatial analysis of
  professional basketball.
\newblock In {\em International conference on machine learning\/} (2014), PMLR,
  pp.~235--243.

\bibitem{morgulev2018sports}
{\sc Morgulev, E., Azar, O.~H., and Lidor, R.}
\newblock Sports analytics and the big-data era.
\newblock {\em International Journal of Data Science and Analytics 5}, 4
  (2018), 213--222.

\bibitem{nishimura2017discontinuous}
{\sc Nishimura, A., Dunson, D., and Lu, J.}
\newblock Discontinuous {Hamiltonian Monte Carlo} for sampling discrete
  parameters.
\newblock {\em arXiv preprint arXiv:1705.08510 853\/} (2017).

\bibitem{nistala2019using}
{\sc Nistala, A., and Guttag, J.}
\newblock {\em Using Deep Learning to Understand Patterns of Player Movement in
  the NBA}.
\newblock MIT Sloan Sports Analytics Conference, Boston, MA, 2019.

\bibitem{reid2016matchplay}
{\sc Reid, M., Morgan, S., and Whiteside, D.}
\newblock Matchplay characteristics of grand slam tennis: implications for
  training and conditioning.
\newblock {\em {Journal of Sports Sciences} 34}, 19 (2016), 1791--1798.

\bibitem{rutherford2017skills}
{\sc Rutherford, J.}
\newblock {\em Skills, Drills \& Strategies for Tennis}.
\newblock New York, NY, 2017.

\bibitem{smith2017absolute}
{\sc Smith, M., and Stolle, F.}
\newblock {\em Absolute Tennis}.
\newblock New Chapter Press, Incorporated, New York, NY, 2017.

\bibitem{teh2006hierarchical}
{\sc Teh, Y.~W., Jordan, M.~I., Beal, M.~J., and Blei, D.~M.}
\newblock Hierarchical dirichlet processes.
\newblock {\em Journal of the American Statistical Association 101}, 476
  (2006), 1566--1581.

\bibitem{vehtari2017practical}
{\sc Vehtari, A., Gelman, A., and Gabry, J.}
\newblock Practical {Bayesian} model evaluation using leave-one-out
  cross-validation and waic.
\newblock {\em {Statistics and Computing} 27}, 5 (2017), 1413--1432.

\bibitem{vehtari2016bayesian}
{\sc Vehtari, A., Mononen, T., Tolvanen, V., Sivula, T., and Winther, O.}
\newblock Bayesian leave-one-out cross-validation approximations for {Gaussian}
  latent variable models.
\newblock {\em The Journal of Machine Learning Research 17}, 1 (2016),
  3581--3618.

\bibitem{wei2015predicting}
{\sc Wei, X., Lucey, P., Morgan, S., Carr, P., Reid, M., and Sridharan, S.}
\newblock Predicting serves in tennis using style priors.
\newblock In {\em Proceedings of the 21th ACM SIGKDD International Conference
  on Knowledge Discovery and Data Mining\/} (2015), pp.~2207--2215.

\bibitem{yue2014learning}
{\sc Yue, Y., Lucey, P., Carr, P., Bialkowski, A., and Matthews, I.}
\newblock Learning fine-grained spatial models for dynamic sports play
  prediction.
\newblock In {\em {IEEE International Conference on Data Mining}\/} (2014),
  IEEE, pp.~670--679.

\end{thebibliography}

\end{document}